\documentclass[aps,pra,twocolumn,superscriptaddress]{revtex4-2}
\usepackage{epsfig}
\usepackage{natbib}
\usepackage{amsmath,amssymb,mathrsfs}
\usepackage{subfigure}
\usepackage{tabularx}
\usepackage{longtable}
\usepackage{amsfonts}
\usepackage{rotating}
\usepackage{bbold}
\usepackage{hhline}
\usepackage{braket}
\usepackage{txfonts, comment}

\usepackage[unicode=true,bookmarks=true,bookmarksnumbered=false,bookmarksopen=false,breaklinks=false,pdfborder={0 0 1},backref=false,colorlinks=true]{hyperref}

\hypersetup{linkcolor=magenta,urlcolor=blue,citecolor=blue,pdfstartview={FitH},unicode=true}

\begin{document}

\title{Adaptive quantum optimization algorithms for programmable atom-cavity systems}

\author{Yuchen Luo}
\affiliation{State Key Laboratory of Surface Physics, Key Laboratory of Micro and Nano Photonic Structures (MOE), and Department of Physics, Fudan University, Shanghai 200433, China}
\affiliation{Shanghai Branch, Hefei National Laboratory, Shanghai 201315, China} 

\author{Xiaopeng Li}
\email{xiaopeng\underline{ }li@fudan.edu.cn}
\affiliation{State Key Laboratory of Surface Physics, Key Laboratory of Micro and Nano Photonic Structures (MOE), and Department of Physics, Fudan University, Shanghai 200433, China} 
\affiliation{Shanghai Qi Zhi Institute, AI Tower, Xuhui District, Shanghai 200232, China}
\affiliation{Shanghai Artificial Intelligence Laboratory, Shanghai 200232, China} 
\affiliation{Shanghai Research Center for Quantum Sciences, Shanghai 201315, China} 

\author{Jian Lin}
\email{jlin17@fudan.edu.cn} 
\affiliation{State Key Laboratory of Surface Physics, Key Laboratory of Micro and Nano Photonic Structures (MOE), and Department of Physics, Fudan University, Shanghai 200433, China}

\date{\today}

\begin{abstract}

Developing quantum algorithms adaptive to specific constraints of near-term devices is an essential step towards practical quantum advantage. In a recent work [\href{https://journals.aps.org/prl/abstract/10.1103/PhysRevLett.131.103601}{Phys. Rev. Lett. \textbf{131}, 103601(2023)}], we show cold atoms in an optical cavity can be built as a universal quantum optimizer with programmable all-to-all interactions, and the effective Hamiltonian for atoms directly encodes number partitioning problems (NPPs). Here, we numerically investigate the performance of quantum annealing (QA) and quantum approximate optimization algorithm (QAOA) to find the solution of NPP that is encoded in the ground state of atomic qubits. 
We find the success probability of the standard QA decays rapidly with the problem size. The optimized annealing path or inhomogeneous driving fields only lead to mild improvement on the success probability. 
Similarly, the standard QAOA always gets trapped in a false local minimum, and there is no significant performance improvement as we increase the depth of the quantum circuit. Inspired by the counterdiabatic driving, we propose an adaptive ansatz of QAOA which releases the parameter freedom of the NPP Hamiltonian to match higher-order counterdiabatic terms. 
Through numerical simulations, we find that our adaptive QAOA can achieve the optimal solution within very small circuit depth. It is thus worth paying the extra optimization cost of additional parameters for improving QAOA performance. Therefore, our adaptive QAOA provides a promising choice for programmable atom-cavity systems to demonstrate competitive computational power within its quantum coherence time. 

\end{abstract}

\maketitle

\section{Introduction}

Quantum optimization is a computing paradigm that utilizes quantum fluctuations to efficiently solve hard combinatorial optimization problems~\cite{Steiglitz2013combinatorial}. Typically, the relevant cost function is encoded in a programmable quantum many-body system~\cite{Lucas2014fip}, and its ground state corresponding to the optimal solution can be prepared through a controlled dynamics of the quantum system following quantum adiabatic~\cite{Nishimori1998pre,farhi2000arxiv,Das2008rmp,Lidar2018rmp} or variational principles~\cite{farhi2014arxiv,Wecker2016pra,Peruzzo2014nc,Cerezo2021nrp}. Prominent examples of such algorithms include QA and QAOA, both of which are promising candidates to provide significant quantum speedup over classical computing~\cite{Albash2018prx,Hastings2021Quantum,farhi2019quantum,Zhou2020prx}. With decades-long efforts devoted to building large-scale programmable quantum devices~\cite{Hauke2020rpp,Altman2021prx}, these two quantum algorithms have already been implemented on various physical platforms, including superconducting qubits~\cite{Johnson2011nature,Harrigan2021np,Willsch2020qip}, trapped ions~\cite{Tobias2016nc,Guido2020pnas,Monroe2021rmp}, neutral atoms~\cite{Ebadi2022science,Dlaska2022prl,Graham2022nature} and photonics systems~\cite{Mohseni2022nrp}. In spite of these achievements, severe limitations on qubit number and coherence time prevented the realization of practical quantum advantage. The interest of the quantum computing community, therefore, have been raised in searching and improving algorithms to obtain useful results on the present noisy intermediate-scale quantum (NISQ) devices~\cite{Preskill2018quantum}. 

The major bottleneck for the performance of QA is vanishing spectral gaps during annealing dynamics resulting in the requirement for overlong run time that generally scales exponentially with the problem size~\cite{Amin2009pra,Altshuler2010pnas}. Various modifications to QA have been proposed to overcome this problem, such as reverse annealing~\cite{Ohkuwa2018pra}, counterdiabatic driving~\cite{Berry2009iop,Campo2013prl,Dries2017pnas} and using inhomogeneous fields~\cite{Susa2018jpsj,Tobias2019prl}. Besides, variational QA has emerged as a more general approach to reach better performance by adjusting annealing path or other controllable factors with the help of a classical optimizer~\cite{Lin2020pra,Schiffer2022prx,Imoto2022iop,Rudi2024prr}. Inspired by the Trotterized version of QA, QAOA is designed to be a variational quantum algorithm with the ansatz consist of alternating cost and mixer layers. Theoretically, the QAOA monotonically improves with increasing layers and succeeds in the limit of infinitely large circuit depth where the adiabatic evolution is recovered~\cite{farhi2014arxiv}. Recent relevant researches indicate that QAOA intrinsically includes certain counterdiabatic effects to compensate the diabatic excitations~\cite{Wurtz2022quantum,Chai2022pra}, and thus is expected to outperform the QA under a finite operation time. However, in practice the performance of QAOA may even  become worse as we increase the circuit depth due to the unavoidable decoherence effects in the present NISQ devices. Meantime, the classical optimization cost also becomes too heavy for a deep quantum circuit because of the infamous barren plateau problem~\cite{Cerezo2021nrp}. Taking into account such practical limitations, 
designing optimal QAOA ansatz that is feasible to realistic quantum systems is of vital importance to the investigation of computation power of quantum optimization~\cite{Kostas2024pr}. 

Our paper focuses on the controllable quantum system of cold atoms in an optical cavity, which have been widely used for quantum simulations of many-body physics~\cite{Ritsch2013rmp,Reiserer2015rmp,Blais2021rmp}. Remarkable experimental progress in recent years have reported the realization of non-local atomic interactions mediated by cavity photons, and the interaction is programmable through manipulating the positions of atoms with optical tweezers~\cite{Davis2019prl,Muniz2020nature,Periwal2021nature,Liu2023prl}. Based on this technique, a recently proposed Raman coupling scheme utilizing four-photon processes leads to an effective Hamiltonian for atoms that naturally encodes NPPs~\cite{Ye2023prl}, a well-known NP-complete problem whose solution is challenging for both exact and approximate methods~\cite{Pedro2010ejor}. In addition, a wide range of hard optimization problems with arbitrary connectivity can be efficiently mapping to the native NPP Hamiltonian by a systematic method, making the atom-cavity system a promising platform for universal quantum optimization. 

To exploit the computational power of the atom-cavity platform, we investigate adaptive and efficient quantum algorithms based on two widely applied approaches, i.e. QA and QAOA, and test the algorithm performance on hard instances of NPP. In addition to the standard QA, we consider two variational variants with optimized annealing path and inhomogeneous driving fields respectively. By numerical simulation of the quantum time evolution, we find that the success probability of the standard QA and the two variational variants with optimized parameters exponentially decreases as the problem size gets larger, and their failure to efficiently find the optimal solution of NPP can be attributed to the exponentially small energy gap and the exponentially many quasi-optimal solutions. As for the case of QAOA, the optimization of the standard ansatz always ends in a false local minimum, and increasing the circuit depth barely produces any improvement on the success probability, and at the same time, the classical optimization becomes much harder. Inspired by counterdiabatic driving~\cite{Berry2009iop,Campo2013prl,Dries2017pnas}, we propose an ansatz adaptive to the atom-cavity platform which releases the parameter freedom of the NPP Hamiltonian to match higher-order counterdiabatic terms. Numerical results show that this adaptive QAOA can find the optimal solution of NPP with much smaller circuit depth compared to the standard QAOA. The extra optimization cost by classical computing is reasonable for the small circuit depth. Our results imply that the adaptive QAOA is a promising algorithm to implement efficient quantum optimization on the programmable atom-cavity platform.

The remainder of this paper is organized as follows. In Sec. \ref{sec:2}, we give the required background on NPP and identify its hard instances to test the algorithms. Sec. \ref{sec:3} presents the numerical results on the performance of the standard QA and our two variational variants, reflecting the hardness of solving NPP by quantum computing. In Sec. \ref{sec:4}, we numerically investigate the performance of the standard QAOA on NPP, and propose an adaptive ansatz of QAOA for the atom-cavity platform as inspired by the idea of counterdiabatic driving. We find significant improvement in using the adaptive QAOA for solving NPP, as compared to the standard QAOA. Finally, we provide concluding remarks in Sec. \ref{sec:5}.

\section{Number partitioning problem}\label{sec:2}

NPP is one of Garey and Johnson’s six basic NP-complete problems which form the core of the theory of NP completeness~\cite{garey1979computers}, and has important applications in areas like public key encryption and task scheduling. It is defined as follows~\cite{Lucas2014fip}: Given a set of positive numbers $\mathcal{A}=\{a_1,a_2,\dots a_N\}$ with $a_i\in \{1,2,\dots,A\}$, find a partition $\mathcal{P}\subset \{1,2,\dots N\}$ that minimizes the difference $E(\mathcal{P})=\left| \sum_{i\in\mathcal{P}} a_i - \sum_{j\in \bar{\mathcal{P}}} a_j\right|$. A partition with $E=0\ (E=1)$ for even (odd) $\sum_{i=1}^N a_i$ is called perfect partition.  

The minimum partition is equivalent to the ground state of an infinite-range spin glass Hamiltonian
\begin{equation}
   \hat{H}_{\text{P}}=\left( \sum_{i=1}^N \tilde{a}_i\hat{\sigma}_i^z\right)^2=\sum_{i,j=1}^N \tilde{a}_i\tilde{a}_j\hat{\sigma}_i^z\hat{\sigma}_j^z,
   \label{eq:H_P}
\end{equation}
where the spin orientation of $\hat{\sigma}_i^z$ being positive (negative) encodes the $i$th integer belonging to $\mathcal{P}\ (\bar{\mathcal{P}})$,  and we bound the interaction strength by rescaling the integers with the number distribution range $\tilde{a}_i=a_i/A$. This Hamiltonian has natural realization with cold atoms in an optical cavity~\cite{Ye2023prl}. 
Coupling the atoms through a photonic cavity mode leads to the desired non-local interactions, and the programmable coupling strength is realized by placing the atoms at different positions in the cavity with optical tweezers. 

The NP-hardness of NPP depends on exponentially large integers in the problem size, i.e. $A=2^{\kappa N}$. As observed in numerous NP-hard problems, the NPP has an ``easy-to-hard" phase transition separating the instances typically having $\mathcal{O}(2^N)$ perfect partitions from the ones with the probability of finding a perfect partition exponentially small~\cite{Mertens1998prl,mertens2005easiest}. There is an abrupt increase in the average computational cost of solving the random instances around the transition point $\kappa_c$. Resorting to the techniques of statistical mechanics, the analytical form of the transition point have been derived as $\kappa_c=1-\frac{\operatorname{log_2}N}{2N} +\mathcal{O}(\frac{1}{N})$. Accordingly, we benchmark the quantum algorithms adaptive to the atom-cavity platform on the random instances with $\kappa=1$ in which the numbers are independent, identically distributed random variables that take on integer values between $1$ and $A=2^N$ with equal probability.

\section{Quantum annealing}\label{sec:3}

\subsection{Algorithm settings}

In the framework of standard QA~\cite{farhi2000arxiv}, the solution to an optimization problem is encoded in the ground state of a problem Hamiltonian $\hat{H}_\text{P}$, and quantum fluctuations to explore the energy landscape of the problem are generated by a driving Hamiltonian $\hat{H}_\text{D}$ that does not commute with $\hat{H}_\text{P}$. The dynamics in the total annealing time $T$ follows a time-dependent Hamiltonian of the form
\begin{equation}
   \hat{H}(t) = \left[1-\lambda(t)\right]\hat{H}_{\text{D}} + \lambda(t) \hat{H}_{\text{P}}, 
   \label{eq:H_QA}
\end{equation} 
which interpolates from $\hat{H}_\text{D}$ to $\hat{H}_\text{P}$ with a linear schedule $\lambda(t)=t/T \in [0,1]$. The driving Hamiltonian is traditionally chosen to be the uniform transverse fields 
\begin{equation}
   \hat{H}_{\text{D}}= -\sum_{i=1}^N \hat{\sigma}_i^x, 
   \label{eq:H_D}
\end{equation}
whose ground state $|+\rangle ^{\otimes N}$ is unique and easy to prepare. Starting from the ground state of $\hat{H}_\text{D}$, the adiabatic theorem of quantum mechanics~\cite{Kato1950jpsj,Messiah1962quantum} ensures the final state $|\psi(T)\rangle$ will converge to the ground state of $\hat{H}_\text{P}$ if the annealing time scales inversely proportional to a polynomial of the minimal energy gap~\cite{Sabine2007jmp,Nishimori2008jmp,Lidar2009jmp}. During the annealing process to solve most combinatorial optimization problems of practical interest, the system undergoes a first-order quantum transition with an exponentially small gap~\cite{Amin2009pra,Altshuler2010pnas}, which makes the problem intractable within the limited coherence time of current NISQ devices.  

Previous studies on QA have indicated that proper choices of the annealing path or the configuration of driving fields may result in dramatically improved performance~\cite{Lin2020pra,Rudi2024prr,Susa2018jpsj}. Based on this conclusion, we consider two variational variants of QA for larger efficiency in a limited annealing time, both of which put no burden of additional terms on practical experiment implementation. 

The first variant aims to find optimal annealing path using the parametrization
\begin{equation}
   \lambda(t)=\frac{t}{T}+\sum_{m=1}^C b_m\operatorname{sin}(\frac{m\pi t}{T}), 
   \label{eq:lambda_t}
\end{equation} 
where $C$ is a cutoff for high frequency components, and $\{b_m\}$ is a set of variational parameters to be determined by a classical optimizer. This parameterization is asymptotically complete as the cutoff $C$ approaches infinity. Intuitively, varying the annealing path slowly in the vicinity of minimum gap can help achieve better performance. 

The second variant keeps the linear form of annealing path, and employs the inhomogeneous transverse driving fields by taking each on-site potential $h_i$ as a variational parameter
\begin{equation}
   \hat{H}_D=-\sum_{i=1}^N h_i\hat{\sigma}_i^x,
   \label{eq:H_D_i}
\end{equation}  
with restriction $h_i>0$ such that the ground state of the driving Hamiltonian remains the uniform superpositions of all possible qubit configurations. 

Both our variational approaches take the cost function to be the expectation value of the problem Hamiltonian at the final state 
\begin{equation}
E(T)=\langle \psi(T) |\hat{H}_{\text{P}} | \psi(T)\rangle,
\label{eq:E_T} 
\end{equation} 
which can be obtained by repeated measurements of the quantum system in the computational basis. 

\begin{figure}[htp]
   \centering
   \includegraphics[width=0.48\textwidth]{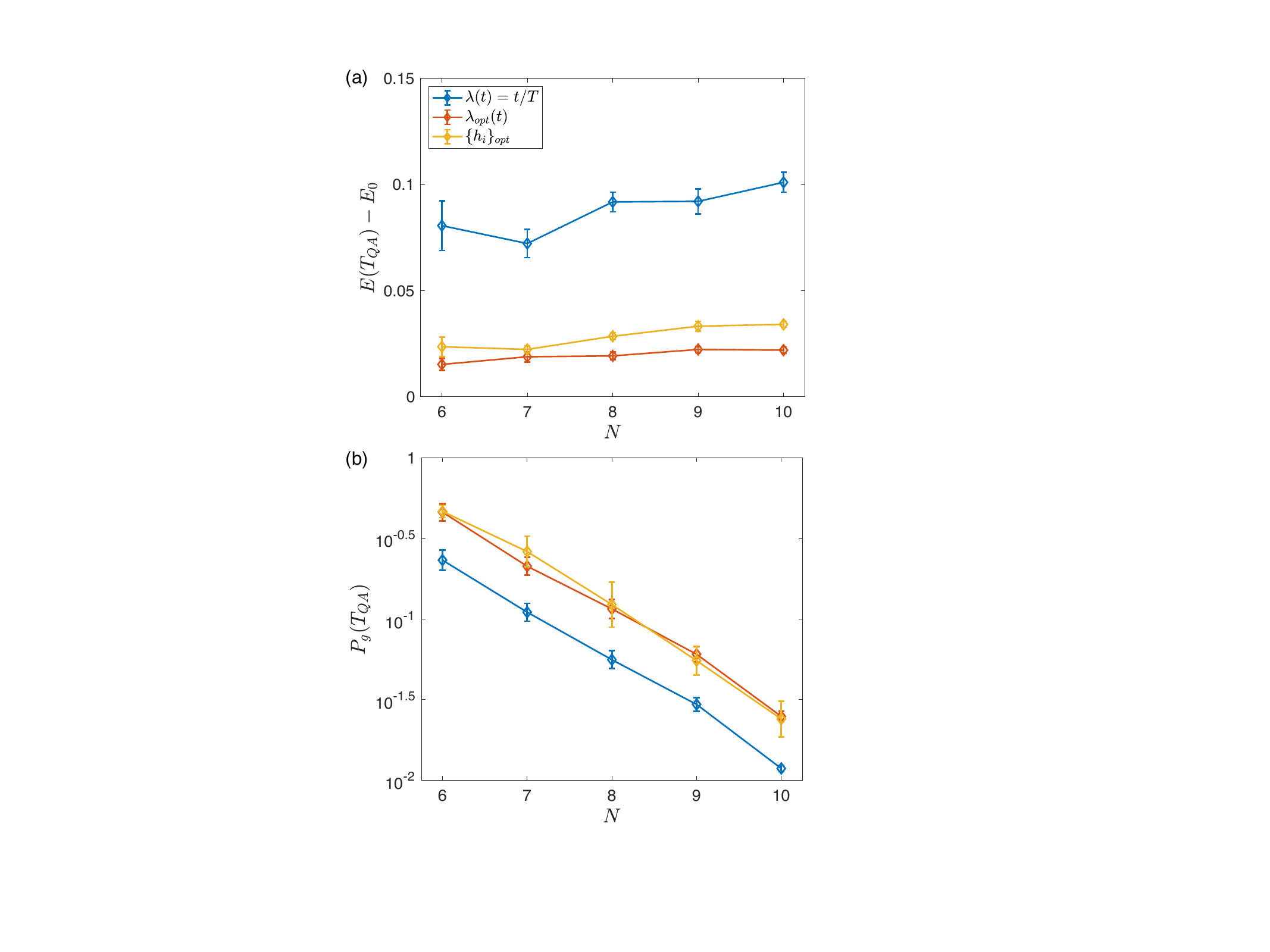}
   \caption{(color online) (a) The approximation error and (b) the success probability of the standard QA and the two variational variants respectively with optimized annealing path and inhomogeneous driving fields as a function of the problem size from $N=6$ to $N=10$. Here, we set the total annealing time $T_{\text{QA}}=50$, which is reasonably accessible to the atom-cavity platform~\cite{Ye2023prl}. The results are averaged over $10$ different NPP random instances for each problem size.
   }
   \label{fig:1}
\end{figure}

\begin{figure}[htp]
   \centering
   \includegraphics[width=0.48\textwidth]{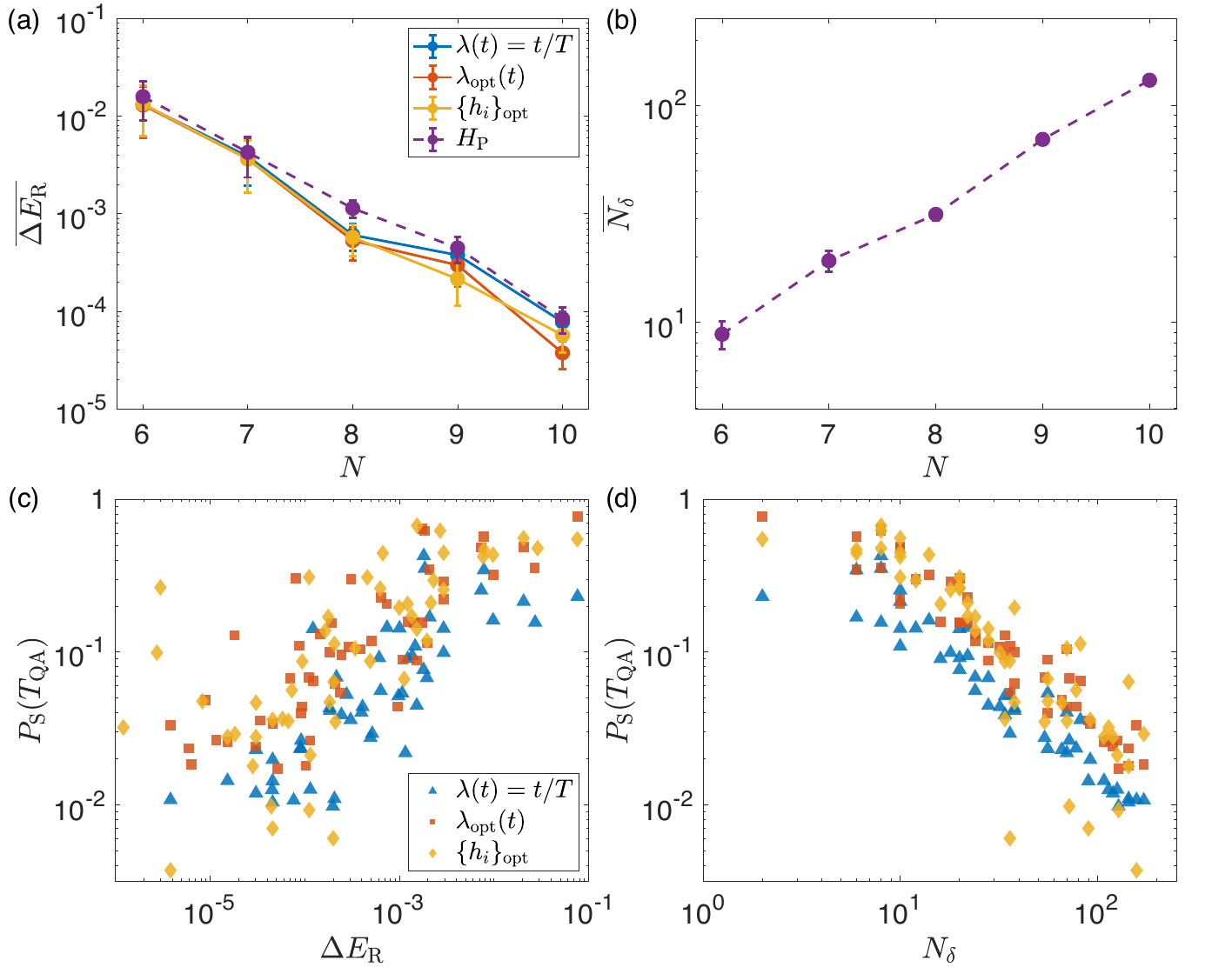}
   \caption{(color online) (a) The average relevant gap for the standard QA and our two variational variants to solve the NPP random instances as a function of the problem size. The average energy gap of the problem Hamiltonian, as depicted with a dashed line, sets a upper bound for the relevant gaps. (b) The number of the quasi-optimal solutions averaged over the NPP random instances as a function of the problem size. The individual success probability for the standard QA and our two variational variants to solve the total $50$ random instances of all the considered problem sizes in the annealing time $T_{\text{QA}}=50$ versus (c) the relevant gap and (d) the number of the quasi-optimal solutions. Each symbol represents the result of one instance. }
   \label{fig:2}
\end{figure}

\subsection{Numerical Results}
\label{sec:3b}

We evaluate the performance of QA by two metrics: the approximation error $\epsilon$ and the success probability $P_{\text{S}}$~\cite{Ebadi2022science}. The former metric $\epsilon$ measures the difference between the final energy and the ground-state energy of the problem Hamiltonian $\hat{H}_{\text{P}}$ by the quantity
\begin{equation}
   \epsilon(T)=1-\frac{E_{\text{max}}-E(T)}{E_{\text{max}}-E_{\text{min}}}, 
   \label{eq:E_res}
\end{equation}
where $E_{\text{max}}$ ($E_{\text{min}}$) is the largest (smallest) eigenvalue of $\hat{H}_{\text{P}}$. The second metric $P_{\text{S}}$ is directly taken as the total probability of finding the optimal solutions in the projective measurements of the final state
\begin{equation}
   P_{\text{S}}(T) = \sum_{i=1}^D |\langle \psi^i_{\text{g}}|\psi(T)\rangle|^2, 
   \label{eq:Pg}
\end{equation}
where $|\psi^i_{\text{g}}\rangle$ is the ground state of the problem Hamiltonian $\hat{H}_{\text{P}}$, and $i=1,\dots,D$ labels the ground-state degeneracy.

To test the efficiency of the quantum algorithms implemented on the programmable atom-cavity platform, we numerically investigate the algorithm performance to solve the NPP random instances with integer numbers distributed in the range $[1,2^N]$, which belong to the hard-to-solve region in the context of classical computing. Taking into account the limitation of computing resources, we consider the problem size $N=6,7,8,9,10$ and generate $10$ different random instances for each size. The total annealing time is set to $T_{\text{QA}}=50$, which is comparable to the quantum coherence time of the atom-cavity system~\cite{Ye2023prl}. For each instance, we simulate the quantum time evolution following the Schrödinger equation, and compute the approximation error $\epsilon(T_{\text{QA}})$, and the success probability $P_{\text{S}}(T_{\text{QA}})$ of the standard QA and our two variational variants with optimized parameters. The implementation of the classical optimization loops to minimize the cost function (\ref{eq:E_T}) is realized by the \texttt{MultiStart} object in the \texttt{Global\  Optimization\ Toolbox} provided by \texttt{MATLAB}, which finds the global optimum through repeatedly running a local optimizer from sufficiently many different initializations. We start the classical optimization from $50$ random initializations for both variational variants, and limit the range of parameters to $-1\leq b_m\leq 1$ and $0< h_i\leq 1$ respectively. The cutoff of the parameterized annealing path (\ref{eq:lambda_t}) in the first variant, i.e. the parameter number, is taken to $C=6$ for all problem sizes, while the parameter number of the second variant with variational driving fields is by definition equal to the problem size $N$.

The performance of the standard QA and our two variational variants averaged over random instances for each problem size is presented in Fig. \ref{fig:1}. As shown in Fig. \ref{fig:1} (a), the optimized annealing path or inhomogeneous driving fields lead to much lower approximation error compared to the standard setting, and counterintuitively there is mild decrease in the approximation error of all the three cases as the problem size gets larger. Despite the decrease in the approximation error, we can see from Fig. \ref{fig:1} (b) that the success probability obtained by all the three approaches exhibits exponentially decrease as the problem size grows. For the considered problem sizes, the two variational variants with optimized parameters have comparable constant improvement on the exponential decay. These results indicate that the standard QA and our two variational variants fail to efficiently find the optimal solution of NPP, and thus are not ideal choices for the atom-cavity system to demonstrate quantum advantage.   

To understand the hardness for the QA to solve NPP, we look into the low-energy spectra of the annealing Hamiltonian (\ref{eq:H_QA}) and the problem Hamiltonian $\hat{H}_\text{P}$. Considering the presence of degeneracy in the problem Hamiltonian, it is reasonable to define a relevant gap 
\begin{equation}
   \Delta E_{\text{R}} = \text{min}_{\lambda\in [0,1]}[E_{D}(\lambda)-E_0(\lambda)],
\end{equation}
which takes the minimal gap between the instantaneous ground state and the first excited state outside the degenerate ground-state subspace. We can see from Fig. \ref{fig:2} (a) that the relevant gap of the standard QA exponentially decreases as the problem size gets larger and the two variants with optimized parameters can not avoid this exponentially small gap. The performance improvement from the variational variants can be attributed to the fine-tuned low-energy spectrum for a specific instance. Actually, the energy gap of the problem Hamiltonian, i.e. $\Delta E(\lambda=1)=E_D(\lambda=1)-E_0(\lambda=1)$, sets an upper bound for the relevant gaps during the whole annealing process. This gap also suffers from the exponential decrease with the problem size, as depicted in Fig. \ref{fig:2} (a) with a dashed line, indicating the intrinsic hardness of NPP. Consequently, there is left little room for performance improvement following the philosophy of adiabaticity. In addition, we count the number of energy levels of the problem Hamiltonian whose energy are higher than its ground-state energy within a small difference $\delta$, and refer to this quantity as the number of the quasi-optimal solutions $N_{\delta}$. The result on $N_{\delta}$ with $\delta=0.1$ is presented in Fig. \ref{fig:2} (b). The exponentially many quasi-optimal solutions with the problem size imply a dense structure of low-energy levels of NPP, which could be a reason for the mild decrease of the approximation error in Fig. \ref{fig:1} (a). In Fig. \ref{fig:2} (c) and (d), we display the individual success probability for our three approaches to solve the total $50$ random instances of all the considered problem sizes in the annealing time $T_{\text{QA}}=50$ versus the relevant gap $\Delta E_{\text{R}}$ and the number of the quasi-optimal solutions $N_{\delta}$ respectively. It can be observed that the success probability is as expected to be positively related with the relevant gap and have negative correlation with the number of the quasi-optimal solutions. We thus attribute the inefficiency of QA in solving NPP to the exponential small gap and the exponential many quasi-optimal solutions.

\section{Quantum approximate optimization algorithm}\label{sec:4}

\subsection{Standard ansatz}

Deriving inspiration from Trotterization of QA, the variational ansatz state of QAOA starts from the ground state $|+\rangle ^{\otimes N}$ of the uniform driving fields $\hat{H}_{\text{D}}= -\sum_{i=1}^N \hat{\sigma}_i^x$, and evolves under the problem Hamiltonian $\hat{H}_{\text{P}}$ and the driving Hamiltonian $\hat{H}_{\text{D}}$ alternatively~\cite{farhi2014arxiv}:  
\begin{equation} 
   |\boldsymbol{\beta}, \boldsymbol{\gamma}\rangle=\prod_{k=1}^p e^{-\mathrm{i} \beta_k H_{\text{D}}} e^{-\mathrm{i} \gamma_k H_{\text{P}}}|+\rangle^{\otimes N},
\end{equation}
where $p$ determines the circuit depth of the variational ansatz, and the evolution durations $\boldsymbol{\beta}=(\beta_1,\dots,\beta_p)$ and $\boldsymbol{\gamma}=(\gamma_1,\dots,\gamma_p)$ are two sets of variational parameters. A classical optimizer is used to search for the optimal parameters $(\boldsymbol{\beta}^*, \boldsymbol{\gamma}^*)$ which minimizes the expectation value of the problem Hamiltonian
\begin{equation} 
   E(\boldsymbol{\beta}, \boldsymbol{\gamma})=\langle \boldsymbol{\beta}, \boldsymbol{\gamma}|\hat{H}_{\text{P}}|\boldsymbol{\beta}, \boldsymbol{\gamma}\rangle. 
   \label{eq:E_QAOA}
\end{equation} 
Then the resulting variational state is expected to have a considerable overlap with the ground state of $\hat{H}_{\text{P}}$ which encodes the optimal solution of the problem. 
  
We take a numerical test on the performance of the standard QAOA applied to NPP, and benchmark on the same set of 10 different random instances with the problem size from $N=6$ to $N=10$ as considered in Sec.~\ref{sec:3b}. For each instance and a given circuit depth $p$, we sample $200$ random initial points from the parameter space, and again choose the \texttt{MultiStart} object in the \texttt{Global\  Optimization\ Toolbox} provided by \texttt{MATLAB} for the optimization of the variational parameters $\boldsymbol{\beta}$ and $\boldsymbol{\gamma}$. Symmetries of the cost function (\ref{eq:E_QAOA}) allow us to restrict the parameter space. Generally, we have the relation $E(\boldsymbol{\beta}, \boldsymbol{\gamma})=E(-\boldsymbol{\beta}, -\boldsymbol{\gamma})$ since both the driving Hamiltonian $\hat{H}_{\text{D}}$ and the problem Hamiltonian $\hat{H}_{\text{P}}$ are real-valued. Owing to the $\mathbb{Z}_2$ symmetry of the NPP Hamiltonian (\ref{eq:H_P}), the exponent $e^{-\mathrm{i}\frac{\pi}{2}\hat{H}_{\text{D}}}=\prod_{i=1}^{N}(-\mathrm{i}\hat{\sigma}_i^x)$ commutes through the variational circuit and cancels with its complex conjugate such that the shift $\beta_i\rightarrow \beta_i + \frac{\pi}{2}$ will have no effect on the cost function. Accordingly, the parameter space can be restricted to $\beta_i \in[0,\frac{\pi}{2}]$. For the efficiency of classical optimization, we artificially limit the other variational parameter to a finite range $\alpha_i \in[0 ,\pi]$.  

\begin{figure*}[htp]
   \centering
   \includegraphics[width=1\textwidth]{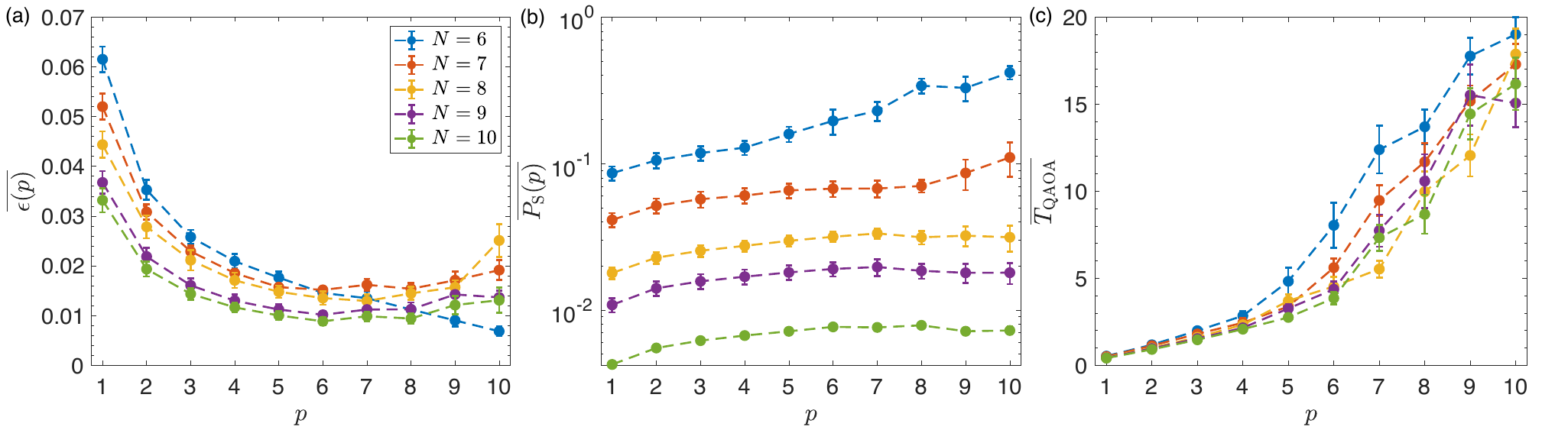}
   \caption{(color online) (a) The approximation error, (b) the success probability and (c) the total evolution duration of the standard QAOA versus the circuit depth for the problem sizes from $N=6$ to $N=10$. The results are averaged over 10 NPP random instances for each problem size.}
   \label{fig:3}
\end{figure*}

\begin{figure*}[htp]
   \centering
   \includegraphics[width=1\textwidth]{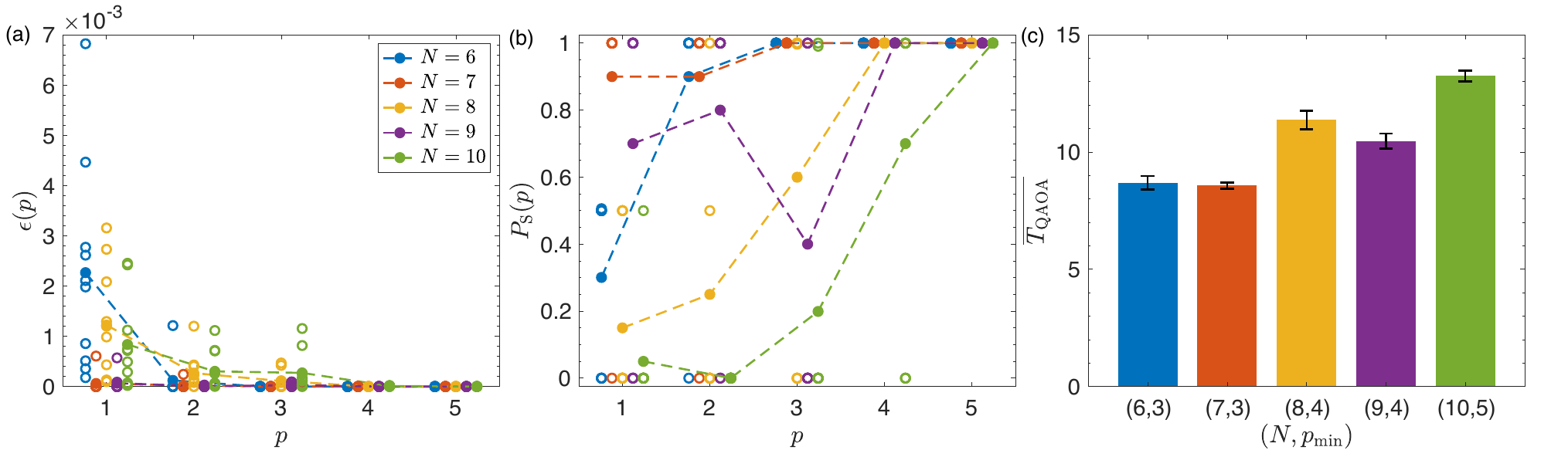}
   \caption{(color online) (a) The approximation error and (b) the success probability of the adaptive QAOA to solve individual instances (empty symbols) and averaged over random instances (solid symbols) versus the circuit depth for the problem sizes from $N=6$ to $N=10$. (c) The total evolution duration corresponding to the minimal circuit depth $p_{\text{min}}$ with which the average success probability for a given problem size N reaches at least 0.99 in the case of the adaptive QAOA.}
   \label{fig:4}
\end{figure*}

The approximation error (\ref{eq:E_res}) and the success probability (\ref{eq:Pg}) are general metrics to evaluate the performance of quantum optimization algorithms. For the case of QAOA, the two metrics are rewritten as $\epsilon(\boldsymbol{\beta},\boldsymbol{\gamma})=1-\frac{E_{\text{max}}-E(\boldsymbol{\beta},\boldsymbol{\gamma})}{E_{\text{max}}-E_{\text{min}}}$ and $P_{\text{S}}(\boldsymbol{\beta},\boldsymbol{\gamma}) = \sum_{i=1}^D |\langle \psi^i_{\text{g}}|\boldsymbol{\beta},\boldsymbol{\gamma}\rangle|^2$ in terms of the circuit parameters. In Fig.~\ref{fig:3} (a) and (b), the two metrics averaged over random instances for each problem size are presented as a function of the circuit depth up to $p=10$. With very shallow circuits, the approximation error takes a rather high value for all the problem sizes, and then displays a downward trend in general as the circuit depth $p$ gets larger. However, except for the case of the smallest problem size $N=6$, increasing the circuit depth to $p=7$ and larger $p$ even leads to higher approximation errors. This behavior can be attributed to the higher difficulty of finding the optimal QAOA parameters for larger problem sizes and larger circuit depth within a limited number of queries to the classical optimizer. As for the success probability, there is no evident increase as the circuit depth grows from $p=1$ to $p=10$, and the value at the circuit depth $p=10$ is still quite low and decays almost exponentially as the problem size becomes larger. The results indicate that the standard QAOA always gets trapped in a false local minimum and increasing the circuit depth for better performance will require notably increasing computing resources on classical optimization.

Out of consideration for practical implementation on the atom-cavity platform, we calculate the total evolution duration of the whole circuit, which is directly defined as the sum of all the duration parameters, 
\begin{equation} 
   T_{\text{QAOA}}=\sum_{i=1}^{p} \beta_i +\gamma_i.
\end{equation} 
As shown in Fig. \ref{fig:3} (c), the total evolution duration grows rapidly as the circuit depth gets larger, and has reached a value that comparable to the quantum coherence time of the atom-cavity platform at the circuit depth $p=10$. These numerical results show that the standard QAOA is inefficient in reaching the optimal solution of NPP. 

\subsection{Counterdiabatic effect in QAOA}

Shortcuts to adiabaticity (STA) is a comprehensive approach to speed up the quantum adiabatic process~\cite{Ruschhaupt2019rmp}, and counterdiabatic driving, also termed transitionless driving, is one of STA techniques that analytically compensates the diabatic transitions by adding an auxiliary driving term to the original time-dependent Hamiltonian~\cite{Berry2009iop}
\begin{equation}
   \hat{H}_{\text{CD}}(t) = \hat{H}(t) + \dot{\lambda}(t)\hat{A}_{\lambda}(t). 
   \label{eq:H_CD}
\end{equation} 
The additional term $\hat{A}_{\lambda}$ is known as the adiabatic gauge potential (AGP)~\cite{Kolodrubetz2017pr}. The exact form of AGP requires knowledge of the spectral properties of the instantaneous Hamiltonians and tends to involve highly non-local couplings, making its implementation a formidable task. A recently proposed method provides a general way to systematically construct the AGP by the nested commutators~\cite{Claeys2019prl}
\begin{equation}
   \begin{aligned}
      \hat{A}_\lambda^{(l)} & =i \sum_{k=1}^l \alpha_k(t) [\underbrace{\hat{H},[\hat{H}, \ldots [\hat{H}}_{2 k-1}, \partial_\lambda H_0]]] \\
      & =i \sum_{k=1}^l \alpha_k(t) \hat{O}_{2 k-1},
   \end{aligned}
   \label{eq:A_l}
\end{equation}
which retrieves the exact AGP in the limit of expansion order $l\to \infty$.  The coefficients $\{\alpha_k\}$ as variational parameters are determined by minimizing the operator distance between the exact AGP and the approximate one~\cite{Dries2017pnas}.  The variational principle is equivalent to minimizing the Hilbert-Schmidt norm of the operator $\hat{G}_\lambda\left(\hat{A}_\lambda^*\right)=\partial_\lambda \hat{H}+i\left[\hat{A}_\lambda^*, \hat{H}\right]$, which can be formulated as 
\begin{equation}
   \left.\frac{\delta S\left(A_\lambda^*\right)}{\delta A_\lambda^*}\right|_{A_\lambda^*=A_\lambda}=0, \quad S\left(A_\lambda^*\right)=\operatorname{Tr}\left[G_\lambda^2\left(A_\lambda^*\right)\right].
\end{equation}

In practice, just the first few orders of the nested commutators (\ref{eq:A_l}) are considered. We expand the approximate ansatz of AGP to the first order, the counterdiabatic Hamiltonian becomes
\begin{equation}   
   \begin{aligned}
      \hat{H}_{\mathrm{CD}}^{(1)}(t) & =\hat{H}(t)+\dot{\lambda}(t) \hat{A}_\lambda^{(1)}(t) \\
      & =[1-\lambda(t)] \hat{H}_D+\lambda(t) \hat{H}_P+i \dot{\lambda}(t) \alpha_1(t)\left[\hat{H}_D, \hat{H}_P\right],
   \end{aligned}
   \label{eq:H_CD_1}
\end{equation}
where the counterdiabatic driving term is simply approximated by the commutator of the driving Hamiltonian $\hat{H}_\text{D}$ and the problem Hamiltonian $\hat{H}_\text{P}$, and the coefficient $\alpha_1$ can be proven to be negative~\cite{Wurtz2022quantum}.

The standard ansatz of QAOA is originally introduced as an analogue of a Trotterized adiabatic evolution~\cite{farhi2014arxiv}. However, there have been researches indicating that QAOA is at least counterdiabatic and has better performance than finite time adiabatic evolution~\cite{Wurtz2022quantum,Chai2022pra}. We can see this from the effective Hamiltonian of the standard QAOA that is constructed by the second-order Baker-Campbell-Hausdorff expansion
\begin{equation} 
   \begin{aligned}
      e^{-\mathrm{i} \beta_k H_D} e^{-\mathrm{i} \gamma_k H_P} & \sim e^{-\mathrm{i} H_{\text {eff }}}, \quad 1 \leqslant k \leqslant p \\
      H_{\text {eff }} & =\beta_k H_D+\gamma_k H_P-\frac{\mathrm{i} \beta_k \gamma_k}{2}\left[H_D, H_P\right]. 
   \end{aligned}
   \label{eq:H_eff}
\end{equation}
Comparing the results of Eqs. (\ref{eq:H_CD_1}) and (\ref{eq:H_eff}), we can find that the first-order Trotter error generated from alternating evolution under $\hat{H}_{\text{P}}$ and $\hat{H}_{\text{D}}$ has the same form and negative sign with the first-order counterdiabatic term, and thus the parameters $(\boldsymbol{\beta}, \boldsymbol{\gamma})$ of QAOA can be quantified as the inclusion of the counterdiabatic driving term to compensate diabatic excitations.

\subsection{Adaptive ansatz}

Inspired by previous studies showing that introducing additional counterdiabatic terms to the QAOA ansatz can lead to greater performance and reduce the circuit depth~\cite{Chai2022pra,Chandarana2022prr,Yao2021prx}, we propose a modification to the standard ansatz of QAOA which adds new variational parameters to enhance the capability to match higher-order counterdiabatic terms, and at the same time keeps its adaptivity to the atom-cavity platform. 

Considering up to two-body interactions, the expansion of the nested commutators (\ref{eq:A_l}) to every order only gives rise to $\hat{\sigma}_i^y\hat{\sigma}_j^z$ terms and no solely local term, which benefits from the single form of $\hat{\sigma}_i^z\hat{\sigma}_j^z$ interactions in the NPP Hamiltonian (\ref{eq:H_P}). The first-order counterdiabatic term $\left[\hat{H}_D, \hat{H}_P\right]$, also the first-order Trotter error generated by alternating evolution under $\hat{H}_{\text{P}}$ and $\hat{H}_{\text{D}}$ in the QAOA ansatz, exactly turns out to be the $\hat{\sigma}_i^y\hat{\sigma}_j^z$ interactions. For this reason, we release the parameter freedom of the interaction strength in the problem Hamiltonian $\hat{H}_{\text{P}}$, i.e. replacing $\hat{H}_{\text{P}}$ that encodes the optimization problem by an arbitrary NPP Hamiltonian 
\begin{equation}
   \hat{H}_{\text{NPP}}(\boldsymbol{\alpha})=\sum_{i,j=1}^N \alpha_i\alpha_j\hat{\sigma}_i^z\hat{\sigma}_j^z,
   \label{eq:H_NPP}
\end{equation}
where $\boldsymbol{\alpha}=(\alpha_1,\dots,\alpha_N)$ is taken as an additional set of variational parameters. Then the QAOA ansatz is modified to
\begin{equation} 
   |\boldsymbol{\beta}, \boldsymbol{\gamma}, \boldsymbol{\alpha}\rangle=\prod_{k=1}^p e^{-\mathrm{i} \beta_k H_{\text{D}}} e^{-\mathrm{i} \gamma_k H_{\text{NPP}}(\boldsymbol{\alpha})}|+\rangle^{\otimes N}.
\end{equation}
Unlike previous studies introducing additional interactions as counterdiabatic terms which can heavily burden the practical implementation, our ansatz just takes additional variational parameters based on the original circuit structure, and the parameterized NPP Hamiltonian $ \hat{H}_{\text{NPP}}(\boldsymbol{\alpha})$ realized on the programmable atom-cavity system is determined by $N$ parameters instead of $\mathcal{O}(N^2)$ for a generic Hamiltonian with two-body interactions. The addition of variational parameters, on one hand, helps the coefficient of the effective $\hat{\sigma}_i^y\hat{\sigma}_j^z$ terms generated from alternating evolution layers to match higher-order counterdiabatic driving; on the other hand, increases the flexibility of the variational ansatz, making it possible to reach larger parts of the Hilbert space with a smaller circuit depth than the standard one. 

To make a fair performance comparison, we take the same set of random instances and the same optimization settings as in the case of the standard QAOA for simulating the performance of the adaptive QAOA to solve NPP. Regarding the $N$ new parameters which determines the interaction strength of the NPP Hamiltonian, we restrict their range to $\alpha_i\in [-0.5,0.5]$ for efficient classical optimization. In Fig. \ref{fig:4} (a) and (b), we show the results of the approximation error and the success probability versus the circuit depth. The results for an individual instance and averaged over random instances are respectively marked by empty and solid symbols, and the different problem sizes are distinguished by different colors. We can see from Fig. \ref{fig:4} (a) that the approximation error for each instance of all the problem sizes at the circuit depth $p=1$ has reached a level which is lower than that of the standard QAOA by one order of magnitude, and vanishes to zero within the circuit depth $p=5$. Correspondingly, the success probability in all the cases, as shown in Fig. \ref{fig:4} (b), converges to unity as the approximation error vanishes. In Fig. \ref{fig:4} (c), we present the total evolution duration corresponding to the minimal circuit depth $p_{\text{min}}$ with which the average success probability for a given problem size $N$ reaches at least 0.99. For all the considered problem size, the minimal evolution duration to obtain the optimal solution is below the quantum coherence time of the atom-cavity platform, and there is no significant growth as the problem size increases. 

\begin{figure}[htp]
   \centering
   \includegraphics[width=0.48\textwidth]{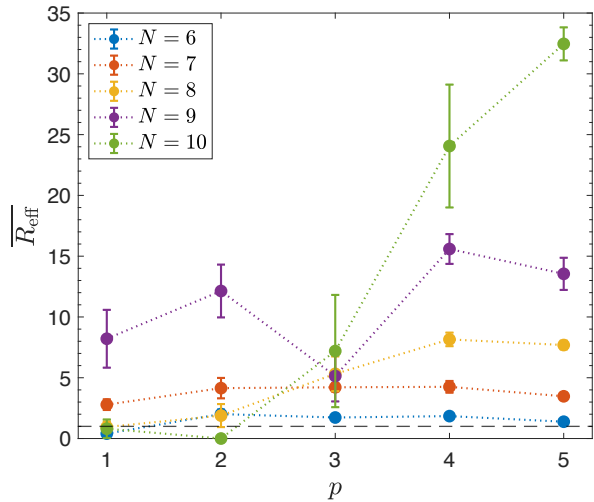}
   \caption{(color online) The optimization efficiency ratio versus the circuit depth for the problem sizes from $N=6$ to $N=10$. The results are averaged over the $10$ random instances for each problem size, and the baseline $R_{\text{eff}}=1$ is drawn with a dashed line which decides whether it is worth paying extra optimization cost in the adaptive QAOA for better performance.}
   \label{fig:5}
\end{figure}

Since the adaptive ansatz has more variational parameters compared to the standard one, the number of cost function evaluations of the adaptive QAOA is supposed to exceed that of the standard QAOA. To decide whether it is worth paying the extra optimization cost for performance improvement, we define the optimization efficiency as the value of the success probability to the number of cost function evaluations
\begin{equation} 
   I_{\text{eff}}=P_{\text{S}}/N_{\text{eval}},
\end{equation} 
and compare the optimization efficiency of the adaptive and the standard QAOA by their ratio
\begin{equation} 
   R_{\text{eff}}=I'_{\text{eff}}/I_{\text{eff}}, 
\end{equation}
where $I'_{\text{eff}}(I_{\text{eff}})$ is the value for the adaptive(standard) QAOA. The optimization efficiency ratio $R_{\text{eff}}>1$ represents that it is deserved to pay the extra optimization cost in the adaptive QAOA for better performance. In Fig. \ref{fig:5}, the average optimization efficiency ratio versus the circuit depth for all the considered problem sizes is depicted, and the baseline $R_{\text{eff}}=1$ is drawn with a dashed line. We can observe that the optimization efficiency ratio is over the baseline for almost all the cases and the optimization efficiency of the adaptive QAOA tends to have a more obvious advantage than that of the standard QAOA as the circuit depth and the problem size get larger. The overall results show the adaptive QAOA is an effective way to significantly improve the performance of QAOA on NPP, and thus a promising quantum algorithm to exploit the computational power of the atom-cavity platform.

\section{Conclusion}
\label{sec:5}

In this work, we aim at finding efficient quantum algorithms adaptive to the programmable atom-cavity system that naturally encodes NPP. We focus on two classes of widely applied algorithms, i.e. QA and QAOA, and numerically investigate their performance on hard instances of NPP. Besides the standard QA, we consider two variants respectively with optimized annealing path and inhomogeneous driving fields for better performance within limited quantum coherence time. Our numerical results show that the success probability of the three QA approaches decreases rapidly with the problem size, implying their failure to solve NPP of the problem size that is practically valuable. To understand the hardness for QA to solve NPP, we look into the low-energy spectra of the annealing Hamiltonian and the problem Hamiltonian, and find that the inefficiency of QA on NPP can be attributed to the exponentially small energy gap and the exponentially many quasi-optimal solutions. In the case of QAOA, the optimization of the standard ansatz always ends in a false local minimum, and increasing the circuit depth only brings minor improvement on the success probability but higher difficulty for classical optimization. The standard QAOA is thus inefficient in reaching the optimal solution of the NPP. In contrast, our proposing adaptive QAOA, which includes higher-order counterdiabatic interaction terms, could reach the optimal solution of NPP within a relatively small quantum circuit accessible to the current atom-cavity technology.

\medskip 

\begin{acknowledgments}
This work is supported by National Program on Key Basic Research Project of China (Grant No. 2021YFA1400900), National Natural Science Foundation of China (Grant No. 11774067, 11934002 and 12304555), Shanghai Municipal Science and Technology Major Project (Grant No. 2019SHZDZX01), Shanghai Rising Star Program (Grant No. 21QA1400500).
\end{acknowledgments}

\bibliography{references}

\end{document}